\documentclass[aps,prl,twocolumn,preprintnumbers,amsmath,amssymb,superscriptaddress]{revtex4}

\usepackage{graphicx}

\usepackage{mathrsfs} 

\begin{document}

\title{Reduced frequency noise in superconducting resonators}

\author{R. Barends}
\author{N. Vercruyssen}
\author{A. Endo}
\affiliation{Kavli Institute of NanoScience, Faculty of Applied
Sciences, Delft University of Technology, Lorentzweg 1, 2628 CJ
Delft, The Netherlands}

\author{P. J. de Visser}
\affiliation{Kavli Institute of NanoScience, Faculty of Applied
Sciences, Delft University of Technology, Lorentzweg 1, 2628 CJ
Delft, The Netherlands}

\affiliation{SRON Netherlands Institute for Space Research,
Sorbonnelaan 2, 3584 CA Utrecht, The Netherlands}

\author{T. Zijlstra}
\author{T. M. Klapwijk}
\affiliation{Kavli Institute of NanoScience, Faculty of Applied
Sciences, Delft University of Technology, Lorentzweg 1, 2628 CJ
Delft, The Netherlands}

\author{J. J. A. Baselmans}
\affiliation{SRON Netherlands Institute for Space Research,
Sorbonnelaan 2, 3584 CA Utrecht, The Netherlands}

\date{\today}

\begin{abstract}
We report a reduction of the frequency noise in coplanar waveguide
superconducting resonators. The reduction of 7 dB is achieved by
removing the exposed dielectric substrate surface from the region
with high electric fields and by using NbTiN. In a model-analysis
the surface of NbTiN is found to be a negligible source of noise,
experimentally supported by a comparison with NbTiN on SiO$_x$
resonators. The reduction is additive to decreasing the noise by
widening the resonators.
\end{abstract}

\maketitle

The development of large and sensitive imaging arrays for far
infrared astronomical instrumentation is rapidly progressing with
microwave kinetic inductance detectors \cite{day}. Arrays have
already been taken to ground-based telescopes
\cite{monfardini,schlaerth}, and readout using frequency domain
multiplexing has been demonstrated \cite{yatesapl}. The frequency
noise in these superconducting resonators is two to three orders of
magnitude above the fundamental limit of generation-recombination
noise. The noise has been conjectured to arise from dipole two-level
systems (TLS) in surface dielectrics \cite{gaoapl2007}, which is
supported by recent experiments: The surface has been shown to be a
dominant source of noise, by measurements on the width scaling by
Gao \emph{et al.} \cite{gaoapl2008-2}. Moreover, we have shown that
introducing dielectrics by covering the resonators with SiO$_x$
leads to an increase in the noise \cite{barendsapl}. Noroozian
\emph{et al.} showed that the noise arises predominantly from the
capacitive portion of the resonator by using a lumped element
capacitor \cite{noroozianaipconf}. Noise reduction can be achieved
by widening the resonator, in essence decreasing the surface to
volume ratio \cite{gaoapl2008-2,noroozianaipconf,barendsieee2009}.
However, the practically limiting source of noise remains to be
identified and reduced.

Here, we show that the noise can be decreased by minimizing the
dielectrics in the resonator itself. The lowest noise is achieved by
using NbTiN deposited on top of a hydrogen passivated substrate as
well as by removing the substrate from the region with the largest
electric fields. The combination of removing the substrate and
widening the resonator leads to a reduction of 9 dB for our
first-generation resonators.

Dipole TLS are known to influence the temperature dependent
permittivity $\epsilon$ \cite{phillips}, and consequently the
resonance frequency. The superposition of permittivity and complex
conductivity ($\sigma_1 - i \sigma_2$) controls the resonance
frequency \cite{barendsapl,gaoapl2008}
\begin{equation}
\label{eq:f0}
\frac{\delta f_0}{f_0} = \frac{\alpha \beta}{4} \frac{\delta \sigma_2}{\sigma_2} - \frac{F}{2} \frac{\delta \epsilon}{\epsilon}
\end{equation}
with $\alpha$ the kinetic inductance fraction and $\beta=1$ for the
thick film and $\beta=2$ for the thin film limit. The filling factor
$F$ \cite{gaoapl2008} takes into account the location of the
dielectric and weighs its contribution to the frequency by the
electric field energy inside our resonator geometry. It is defined
by $F={ \frac{1}{2} \epsilon_0 \epsilon_{h} \iiint_{V_{h}}
|\vec{E}(\vec{r})|^2 d\vec{r}} / {\frac{1}{4} C V_r^2 l}$, with $C$
the capacitance per unit length, $V_r$ the standing wave voltage,
$l$ the length of the resonator, and $\epsilon_h$ the relative
permittivity and $V_h$ the volume of the dielectric hosting the TLS.

Similarly, dipole TLS cause frequency noise through the time-varying
permittivity $\epsilon(\vec{r},t)$ \cite{gaoapl2007}. Consequently,
the power spectral density of the permittivity $S_{\epsilon} = 2
\epsilon_0^2 \mathscr{F} \{ \left< \epsilon_h(t) \epsilon_h(t-\tau)
\right> \}$ translates to frequency noise \cite{gaoapl2008-2},
\begin{equation}
\label{eq:sf}
\frac{S_{f_0}}{f_0^2} = \frac{ {\frac{1}{4}}^2 \iiint_{V_{h}}
S_{\epsilon} |\vec{E}(\vec{r})|^4 d\vec{r}} {(\frac{1}{4} C V_r^2 l)^2}
\end{equation}
with $S_{f_0}/f_0^2$ the normalized frequency noise.

\begin{figure}[!b]
    \centering
    \includegraphics[width=1\linewidth]{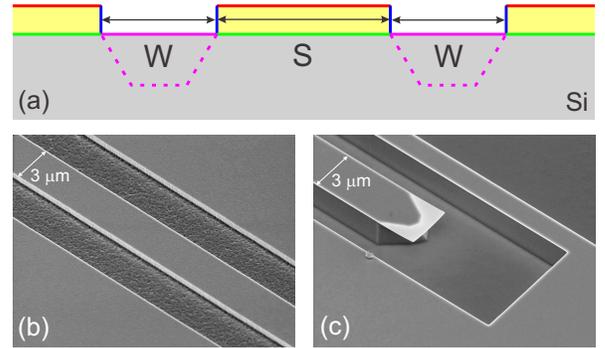}
    \caption{(Color online) (a) The coplanar waveguide with the top metal surface, the exposed substrate,
    the substrate-metal interface and the etched metal edges outlined. (b) A scanning electron microscope image from the standard
    resonator design. (c) The etched grooves near the open end of the resonator, its cross section is outlined in (a) (dashed).
    The central line width is S = 3 $\mu$m in both images.}
    \label{fig:cpw}
\end{figure}

In order to identify the contribution of the various surfaces to
noise we calculate the effect of a hypothetical surface layer with
thickness $t \rightarrow 0$ containing dipole TLS. The electric
fields in the coplanar waveguide geometry are calculated using the
potential matrix to find the charge density. The approach is
detailed in Ref. \cite{barendsarxiv}. We adopt the assumption by Gao
\emph{et al.} \cite{gaoapl2008-2} that the noise spectral density
follows: $S_{\epsilon} = \epsilon_0^2 \kappa /\sqrt{|E|^2+{E_s}^2}$,
with $E_s$ the saturation electric field strength, following the
saturation of microwave loss due to TLS at high intensities
\cite{phillips}.

\begin{figure}[!t]
    \centering
    \includegraphics[width=1\linewidth]{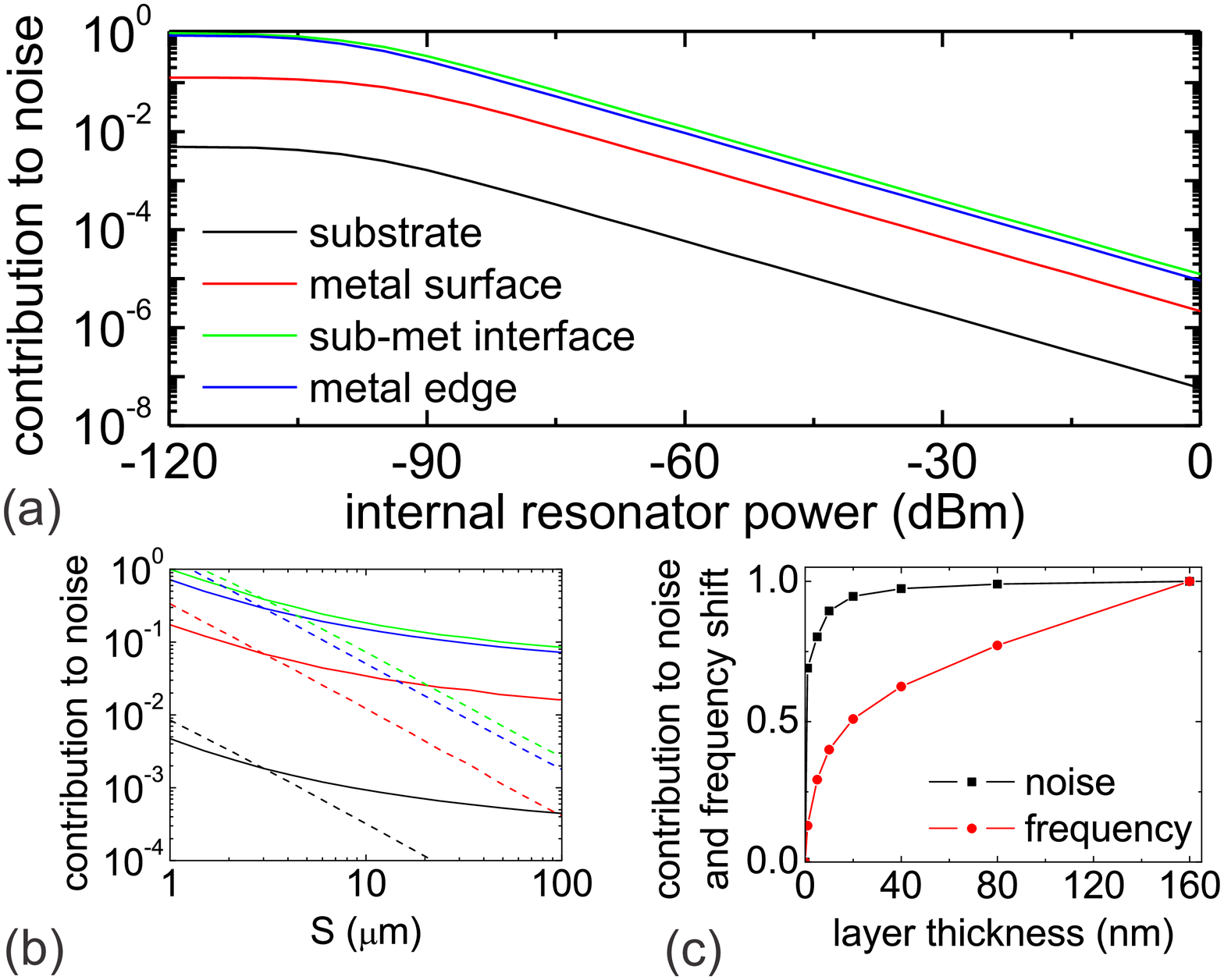}
    \caption{(Color online) (a) The power dependence of the normalized contribution to noise using Eq. \ref{eq:sf}, for a TLS distribution placed on the
    exposed substrate surface, top metal surface, substrate-metal (sub-met) interface and etched metal edges ($E_s=5$ kV/m \cite{barendsarxiv}).
    (b) The normalized contribution to noise of the dielectric layers versus central line width $S$,
    for $W=2$ $\mu$m (solid) and $W=\frac{2}{3}S$ (dashed).
    (c) The normalized contribution to noise and frequency shift for a dielectric layer with finite thickness on top of the metal.
    Calculations are done for $\epsilon_h=1$.}
    \label{fig:noisecalc}
\end{figure}

The hypothetical layer is placed along each of the outlined surfaces
in Fig. \ref{fig:cpw}a, the exposed substrate surface, the top metal
surface, the etched metal edges and the substrate-metal interface.
The results are shown in Fig. \ref{fig:noisecalc}a-b. Importantly,
the contribution to the frequency noise is about two orders of
magnitude larger when the layer is placed on surfaces adjacent to
the metal than when placed on the exposed substrate. This is due to
the high electric fields close to the metal. Moreover, we find that
the noise follows $S_{f_0}/f_0^2 \propto 1/P_{int}^{0.5}$ in the
relevant power range, with $P_{int}$ the internal resonator power
\cite{barendsthesis}. In addition, when widening the resonator
geometry the noise decreases. We have also calculated the influence
of a metal surface dielectric with finite thickness $t$ on the noise
as well as the frequency shift, i.e. Eq. \ref{eq:f0} and Eq.
\ref{eq:sf}. We find that for the noise, only the first few
nanometers matter, whereas the full volume influences the frequency
shift. This is consistent with our previous experiments, where we
showed that frequency deviations arise from the bulk of the
dielectric while noise arises predominantly at surfaces and
interfaces \cite{barendsapl}. The difference arises from the surface
layer being weighed by $|\vec{E}(\vec{r})|^4$ for the noise and
$|\vec{E}(\vec{r})|^2$ for the frequency shift. Interestingly, the
power and width dependence is very similar for each of the surfaces,
and identification of the dominant noise source can be done only by
removing or altering a specific surface.

In order to identify and reduce the dominant noise source, we have
fabricated a series of devices aimed at addressing a specific
surface, see Fig. \ref{fig:cpw}. We use NbTiN quarterwave coplanar
waveguide resonators \cite{day,barendsapl} with varying geometry or
composition. Resonance frequencies lie between 3-5 GHz. As a
reference a 300 nm NbTiN film is DC sputtered on an HF-cleaned high
resistivity ($>1$ k$\Omega$cm) $\left<100\right>$-oriented Si wafer.
Patterning is done using SF$_6$/O$_2$ reactive ion etching. The
critical temperature is $T_c$=14.8 K, the low temperature
resistivity is $\rho$=170 $\mu\Omega$cm and the residual resistance
ratio is 0.94. To identify the importance of using hydrogen
passivated Si, a 300 nm NbTiN film has been deposited on the native
oxide of Si ($T_c=15.5$ K, $\rho=84$ $\mu\Omega$cm and $RRR = 1.0$).
We have also removed the exposed substrate surface from the region
with large electric fields: fully straight, 50 nm thick, NbTiN
resonators are made on Si, aligned along the $\left<110\right>$ axis
of the Si substrate ($T_c=13.6$ K, $\rho=142$ $\mu\Omega$cm and $RRR
= 0.96$). Using KOH wet etching, 0.9 $\mu$m deep grooves (dashed
lines in Fig. \ref{fig:cpw}a) are etched in the gaps along the full
length of the resonators. As a reference for the latter sample as
well as to clarify the influence of the metal edges, a straight, 50
nm thick NbTiN resonator is made where the Si substrate is not
removed.

\begin{figure}[!t]
    \centering
    \includegraphics[width=1\linewidth]{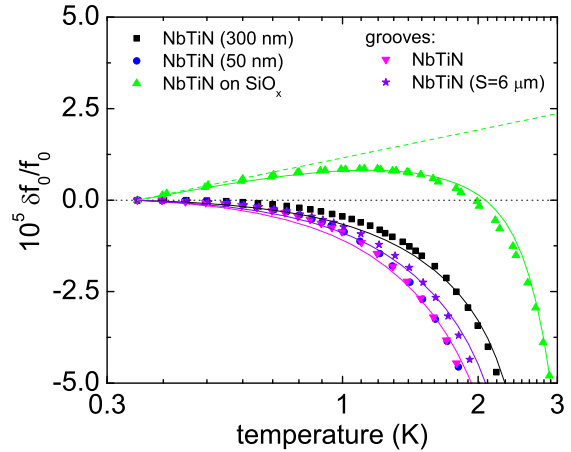}
    \caption{(Color online) The temperature dependence of the fractional resonance frequency. The solid lines are fits using Eq. \ref{eq:f0},
    using Mattis-Bardeen. The NbTiN on SiO$_x$ data follow a superposition of a
    logarithmically temperature dependent permittivity (dashed line) and Mattis-Bardeen (Eq. \ref{eq:f0}).
    We choose $T_0=350$ mK.}
    \label{fig:frequency}
\end{figure}

The frequency noise is measured using a homodyne detection scheme
based on quadrature mixing \cite{day,barendsthesis,barendsapl}. The
samples are cooled to a temperature of 310 mK using a He-3 sorption
cooler placed in a 4.2 K liquid He cryostat. The sample stage is
magnetically shielded with a superconducting shield. We use a low
noise high electron mobility transistor amplifier with a noise
temperature of 4 K \cite{wadefalk}.

The temperature dependence of the resonance frequency is shown in
Fig. \ref{fig:frequency}. For the NbTiN on SiO$_x$ resonator we find
a clear nonmonotonicity. The superposition (Eq. \ref{eq:f0}) of the
complex conductivity and logarithmically temperature dependent
permittivity describes the data. The logarithmic dependence is
consistent with resonant interaction of TLS with the electric fields
at $kT>hf$ \cite{phillips}: $\delta \epsilon/\epsilon = - \ln
\left(T/T_0\right) 2 N p^2/\epsilon$ with $N$ the TLS density of
states, $p$ the dipole moment and $T_0$ an arbitrary reference
temperature. The temperature dependence of the other resonators
follows Mattis-Bardeen \cite{mattis}, provided a broadening
parameter \cite{dynes} of $\Gamma=15-20$ $\mu$eV is included in the
density of states \cite{lambda}.

\begin{figure}[!t]
    \centering
    \includegraphics[width=1\linewidth]{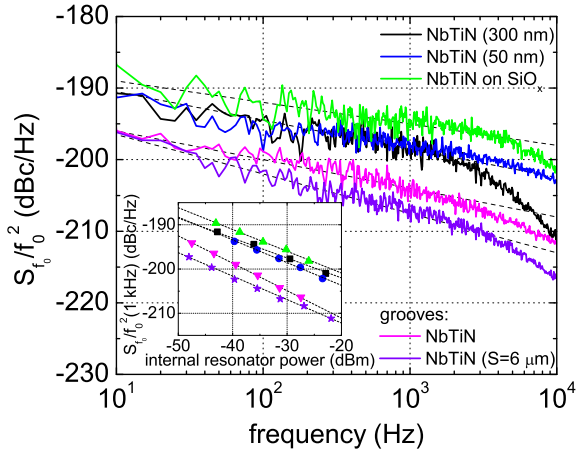}
    \caption{(Color online) The normalized frequency noise spectra of NbTiN, NbTiN on SiO$_x$ and NbTiN with grooves etched in the gaps.
    The widths are $S=3$ $\mu$m, $W=2$ $\mu$m except for the wide resonator with grooves: $S=6$ $\mu$m, $W=2$.
    The bath temperature is 310 mK and the internal resonator power is $P_{int} \approx -30$ dBm.
    The inset shows the power dependence, see Fig. \ref{fig:frequency} for the legend. Dashed lines are fits to the spectral shape and power dependence.}
    \label{fig:noise}
\end{figure}

The frequency noise spectra are shown in Fig. \ref{fig:noise}. The
inset shows the dependence on the internal resonator power
$P_{int}$. The noise spectra follow $S_{f_0}/f_0^2 \propto
1/f^{0.3-0.6}$, until a roll-off frequency on the order of 10 kHz.
This roll-off arises from the resonator-specific response time, set
by the loaded quality factor and resonator frequency. The NbTiN on
SiO$_x$ resonator has the highest frequency noise, at $P_{int} =
-30$ dBm: $S_{f_0}/f_0^2$~(1~kHz)~$=-195$~dBc/Hz. This is 3~dB
noisier than the standard, 300~nm thick, NbTiN resonator which has a
noise level of $-198$~dBc/Hz. The noise remains at $-198$~dBc/Hz
when decreasing the metal thickness by a factor of 6 (8~dB), from
300~nm to 50~nm. Clearly, the noise is decreased with 7~dB for the
resonator with grooves in the gaps, having
$S_{f_0}/f_0^2$~(1~kHz)~$=-205$~dBc/Hz. Moreover, the noise is
reduced over the whole range of spectral frequency and internal
resonator power. For the wider resonator with $S=6$ $\mu$m the noise
is 2~dB lower at $-207$~dBc/Hz. We find that the frequency noise
follows $S_{f_0}/f_0^2 \propto 1/P_{int}^{0.4-0.6}$, up to powers of
$P_{int} \sim -25$~dBm.

The data and analysis show that NbTiN is a clean material and point
towards SiO$_x$ as the dominant source of noise. First, the exposed
Si surface dominates the noise as its removal decreases the noise
considerably. Second, when placing SiO$_x$ below or on top of NbTiN
the noise increases (Fig. \ref{fig:noise} and Ref.
\cite{barendsapl}). Third, the analysis in Fig. \ref{fig:noisecalc}
indicates that the NbTiN surface is clean compared to that of Si, as
the metal surfaces influence the noise much stronger than the
exposed substrate. In addition, the monotonic temperature dependence
of the resonance frequency down to 350 mK indicates that NbTiN has a
minimal dielectric layer, in contrast to Nb, Ta and Al
\cite{barendsapl} as well as NbTiN on SiO$_x$. Moreover, the metal
edges are not dominant, as the noise level is independent of the
thickness of the metal. Finally, the removal of dielectrics from the
gaps leads to a decrease in the capacitance $C$ in Eq. \ref{eq:sf}.
Hence, if the metal surfaces would dominate, the noise would
\emph{increase}. Quantitatively, we estimate $\kappa$(1
kHz)~$\approx 5 \cdot 10^{-27}$ 1/Hz for SiO$_x$, assuming $t=3$ nm.
Importantly, the noise reduction is significant: it is 7 dB below
our standard NbTiN on Si resonators and 11 dB below the lowest
values reported for coplanar waveguide resonators by Gao \emph{et
al.} \cite{gaoapl2007}. In addition, the noise is 2 dB lower when
widening to $S=6$ $\mu$m, which is consistent with our calculation
(1.9 dB, see Fig. \ref{fig:noisecalc}b) and shows that further
improvements can be obtained by widening the resonator.

The data in Fig. \ref{fig:noise} provide a clear guide to low noise
superconducting resonators, by using NbTiN and removing the exposed
substrate surface from the region with the largest electric fields.
A particular approach to remove the dielectric was followed in Ref.
\cite{mazin}. Importantly, we show that both the removal of
dielectrics as well as the widening of the resonator leads to a
significant decrease of the noise. Hence, both approaches can be
considered to be additive to decreasing the noise. Our approach can
be implemented for lumped element resonators
\cite{doyle,noroozianaipconf} as well: by using a
$\left<100\right>$-oriented Si wafer and aligning the fingers and
edges of the interdigitated capacitor along the two perpendicular
$\left<110\right>$ axes, grooves can be etched with a minimal amount
of undercut. Interestingly, our resonators with grooves etched in
the gaps also have higher quality factors at high internal power
levels as well as at the single microwave photon levels needed for
circuit quantum electrodynamics \cite{barendsarxiv}.

To conclude, we have reduced the frequency noise by using NbTiN and
removing the substrate from the region with the highest electric
fields. This indicates that the exposed Si substrate surface is the
main source of the noise, hence the contribution to noise from the
NbTiN surface is not dominant. The followed approach is a
straightforward route to low frequency noise in superconducting
resonators.

\begin{acknowledgments}
The authors thank J. M. Martinis for stimulating discussions. The
work was supported by the Pieter Langerhuizen Lambertuszoon funds of
the Royal Holland Society of Sciences and Humanities and by the EU
NanoSciERA project ``Nanofridge''.
\end{acknowledgments}


\begin{thebibliography}{99}
\bibitem{lambda} For 300~nm thick NbTiN, NbTiN on SiO$_x$, and 50~nm
    thick NbTiN we find for $\alpha \beta/4$: 0.18, 0.08 and 0.33,
    respectively. We infer a magnetic penetration depth $\lambda$ of
    400, 240 and 270 nm.
\end{thebibliography}
\end{document}